\documentclass[review,12pt]{elsarticle}

\usepackage{amssymb}
\usepackage{amsthm}
\usepackage{amsmath}

\usepackage{mathrsfs}
\usepackage{graphicx}
\usepackage{epstopdf}
\usepackage{float}
\usepackage{caption}
\usepackage{subcaption}
\usepackage{bm}
\usepackage{bbm}
\usepackage{mathrsfs}
\usepackage{cleveref}
\usepackage{soul}
\usepackage{accents}
\usepackage{graphicx}
\usepackage{xcolor}
\usepackage{courier} 
\usepackage{listings} 
\usepackage{tabu} 
\usepackage{longtable}
\usepackage{changepage} 
\usepackage[margin=2.2cm]{geometry}
\usepackage{booktabs}

\biboptions{numbers,sort&compress}

\journal{Journal of Power Sources}

\makeatletter
\def\@author#1{\g@addto@macro\elsauthors{\normalsize%
    \def\baselinestretch{1}%
    \upshape\authorsep#1\unskip\textsuperscript{%
      \ifx\@fnmark\@empty\else\unskip\sep\@fnmark\let\sep=,\fi
      \ifx\@corref\@empty\else\unskip\sep\@corref\let\sep=,\fi
      }%
    \def\authorsep{\unskip,\space}%
    \global\let\@fnmark\@empty
    \global\let\@corref\@empty  
    \global\let\sep\@empty}%
    \@eadauthor={#1}
}
\makeatother

\begin{document}

\begin{frontmatter}



\title{Influence of concentration-dependent material properties on the fracture and debonding of electrode particles with core-shell structure}


\author[IC]{Yang Tu}

\author[IC1,FI]{Billy Wu}

\author[SE]{Weilong Ai}

\author[Oxf,IC]{Emilio Mart\'{\i}nez-Pa\~neda\corref{cor1}}
\ead{emilio.martinez-paneda@eng.ox.ac.uk}

\address[IC]{Department of Civil and Environmental Engineering, Imperial College London, London SW7 2AZ, UK}
\address[IC1]{Dyson School of Design Engineering, Imperial College London, London SW7 2AZ, UK}
\address[FI]{The Faraday Institution, Quad One, Becquerel Avenue, Harwell Campus, Didcot, OX11 0RA, UK}
\address[SE]{Jiangsu Key Laboratory of Engineering Mechanics Analysis, School of Civil Engineering, Southeast University, Nanjing, 211189, China}
\address[Oxf]{Department of Engineering Science, University of Oxford, Oxford OX1 3PJ, UK}

\cortext[cor1]{Corresponding author.}

\begin{abstract}
Core-shell electrode particle designs offer a route to improved lithium-ion battery performance. However, they are susceptible to mechanical damage such as fracture and debonding, which can significantly reduce their lifetime. Using a coupled finite element model, we explore the impacts of diffusion-induced stresses on the failure mechanisms of an exemplar system with an NMC811 core and an NMC111 shell. In particular, we systematically compare the implications of assuming constant material properties against using Li concentration-dependent diffusion coefficient and partial molar volume. With constant material properties, our results show that smaller cores with thinner shells avoid debonding and fracture regimes. When factoring in a concentration-dependent partial molar volume, the maximum values of tensile hoop stress in the shell are found to be significantly lower than those predicted with constant properties, reducing the likelihood of fracture. Furthermore, with a concentration-dependent diffusion coefficient, significant barriers to full electrode utilisation are observed due to reduced lithium mobility at high states of lithiation. This provides a possible explanation for the reduced accessible capacity observed in experiments. Shell thickness is found to be the dominant factor in precluding structural integrity once the concentration dependency is accounted for. These findings shed new light on the performance and effective design of core-shell electrode particles.\\  

\end{abstract}

\begin{keyword}

Lithium-ion batteries \sep surface coating \sep core-shell particles \sep NMC \sep fracture \sep multi-physics modeling



\end{keyword}

\end{frontmatter}


\section{Introduction}
\label{Introduction}

\begin{figure}[ht]
    \centering
    \includegraphics[width=14cm]{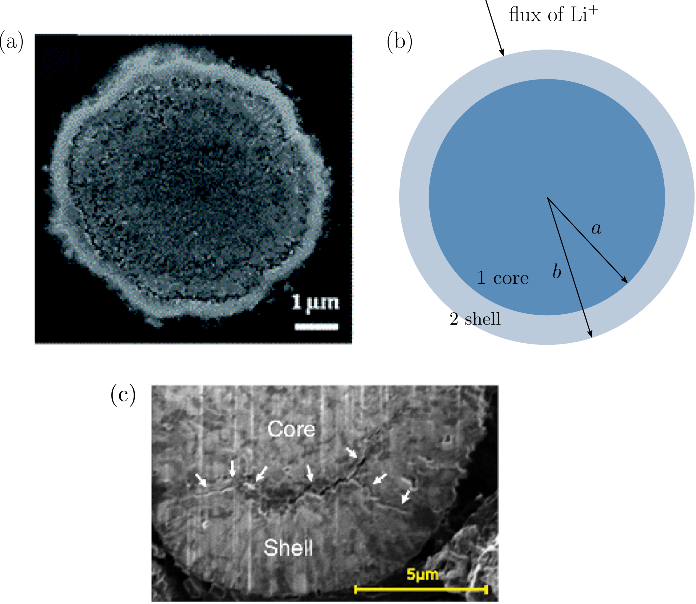}
    \caption{Core-shell design: (a) SEM cross-sectional image of $\mathrm{LiMn_{0.85}Fe_{0.15}PO_4}$ particle coated with $\mathrm{LiFePO_4}$ adapted from \cite{oh2012double}; (b) Schema of an ideal spherical particle with core-shell structure; (c) High-resolution FIB image depicting a cross-section of an NMC core-shell particle after a single charge, revealing shell fracture and core-shell debonding \cite{brandt2020synchrotron}.}
    \label{fig:cs}
\end{figure}

\begin{figure}[H]
    \centering
    \includegraphics[width=14cm]{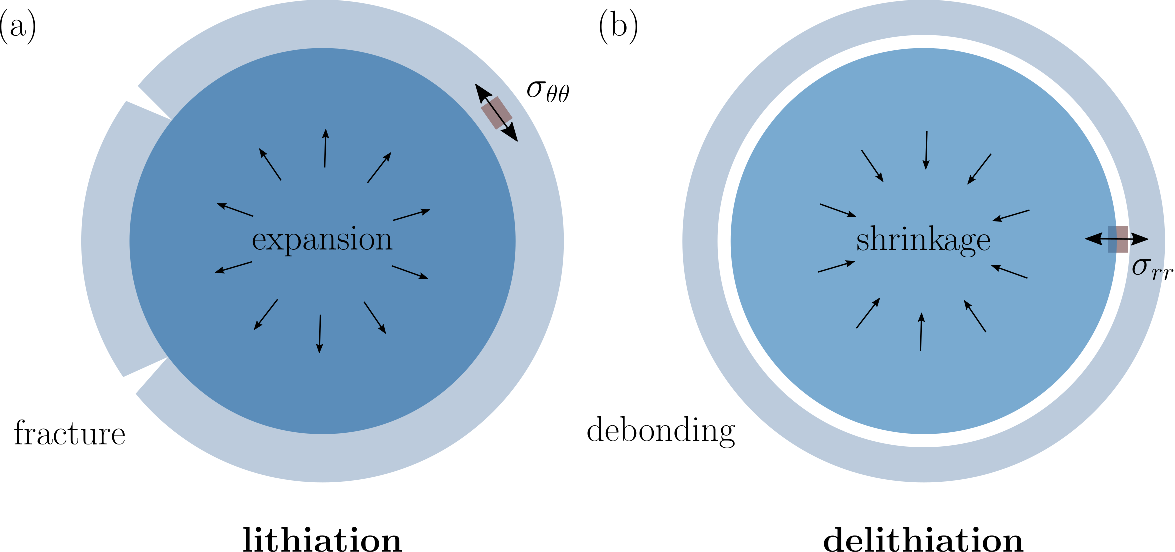}
    \caption{Two failure modes of the core-shell structure: (a) fracture of the shell caused by large hoop stress during lithiation; (b) debonding at the core-shell interface induced by large radial stress during delithiation.}
    \label{fig:failure}
\end{figure}

In recent years, lithium-ion batteries have become a key component in powering an ever-increasing number of portable electronics, electric vehicles, and renewable energy storage systems \cite{li201830}. As the demand for these technologies continues to grow, there is an increasing need to enhance battery performance, extend their lifetime, ensure safety, and reduce costs. One promising approach, which can improve the capacity retention, rate capability, and thermal stability of a lithium-ion battery, is the use of electrode particles with core-shell structures. Electrode particles with surface coatings, as shown in the scanning electron microscopy (SEM) cross-sectional image and schema of Figs. \ref{fig:cs}(a)(b), have demonstrated promising electrochemical performance. For example, 18650 cells with  $\mathrm{LiNi_{0.82}Co_{0.12}Mn_{0.06}O_2}$ cathode particles coated with $\mathrm{LiFePO_4}$ exhibited outstanding cycling stability, with capacity retention of 91.65\% after 500 cycles at 1 C, much higher than that obtained without a surface coating (70.65\%) \cite{zhong2020nano}. The shell can act as a protection barrier that hinders the side reactions between electrode materials and electrolytes \cite{li2022b}. Additionally, with the right choice of materials, the shell can enhance conductivity \cite{li2006cathode}, perform surface modification \cite{fu2006surface}, and mitigate transition metal dissolution \cite{zhao2017improving} of the active materials. However, as shown in Fig. \ref{fig:cs}(c), the shell is prone to fracture and debonding \cite{chen2010,brandt2020synchrotron}, leading to battery degradation. Shell fracture and core-shell debonding are the result of diffusion-induced stresses that arise as a consequence of the lithium intercalation and deintercalation processes, as illustrated in Fig. \ref{fig:failure}.\\ 

The use of coupled, multi-physics modelling has been proven to be an effective approach to better understand the interplay between mechanical and electrochemical effects \cite{Zhao2012,Ma2015,Ai2022,Boyce2022,JMPS2022b}. Diffusion-induced stress (DIS) models have been extensively used for predicting the mechanical response of lithium-ion battery electrodes \cite{christensen2006stress,zhang2007numerical,Zhao2019}. In recent years, DIS models have been extended to gain insight into the coupled mechanical and electrochemical behaviours of particles with core-shell structure. For instance, Hao and Fang \cite{Hao2013} extended the DIS model to spherical core-shell particles, while Zhao et al. \cite{Zhao2012} proposed a dimensional analysis to calculate the energy release rates of shell fracture and debonding based on mechanical stress magnitudes. Building upon these previous works, Wu and Lu \cite{Wu2017} applied a more physically rigorous interface condition that assumes a continuous chemical potential across the core-shell interface. Based on the DIS model, two failure modes of the core-shell structure can be predicted: shell fracture and debonding, which are depicted in Fig. \ref{fig:failure}.\\

Modelling efforts typically assume constant (time- and space-independent) material properties, but recent studies have shown that the diffusion coefficient of Nickel Manganese Cobalt oxide (NMC), one of the most popular cathode materials, can vary by three orders of magnitude during charging \cite{chen2020}. Experiments also revealed that the volume change of the NMC materials is not linear with the amount of lithium inserted \cite{Biasi2017}. During the charging process, Ni-rich NMC materials undergo a phase transition, shifting from the pristine hexagonal H1 phase to two subsequent hexagonal phases, H2 and H3 \cite{li1993situ,Zheng2019}. As lithium gradually departs from the lattice, it causes changes in lattice parameters and chemical bonding between the lithium and transition metals (TM) \cite{dixit2017origin}. These changes affect macroscopic properties such as diffusion coefficient and partial molar volume \cite{Koerver2018}. The evolution of the diffusion coefficient during charging or discharging can be captured by experimental measurements including galvanostatic or potentiostatic intermittent titration techniques (GITT or PITT, respectively) and depolarization. Changes in diffusivity can also be predicted by \textit{ab initio} calculations \cite{wang2022review}. The diffusion coefficient primarily affects the concentration gradient, thereby affecting diffusion-induced stresses. Meanwhile, the partial molar volume directly influences volume changes and, consequently, diffusion-induced stresses. The concentration-dependent partial molar volume can be determined by measuring lattice parameters through \textit{in situ} X-ray diffraction \cite{Biasi2017}, and can also be estimated through \textit{ab initio} methods, such as density function theory (DFT) calculations \cite{Woodcox2021}. The high sensitivity of diffusivity and partial molar volume to Li-ion concentration is consistently observed in experiments across various relevant electrode materials. In Table \ref{table:1}, we gather a list of experimental and first principles studies measuring changes in diffusion coefficient and partial molar volume for a wide range of materials and techniques. Hence, there is a need to incorporate this concentration-dependency into modelling efforts, and to quantify its impact on design rules and failure predictions. This is the objective of this work.

\begin{table}[H]
\centering
\begin{tabular}{cccc}
\toprule  
Material property&Measurement/&Material\\
&calculation technique&\\
\midrule 
& GITT & NMC \cite{chen2020,liebig2019parameterization,Gaoy2020,Wu2012,zhu2022modulating}, LFP \cite{han2004electrochemical,ren2022fe}, LCO \cite{hess2015determination, choi1995effects}, \\
& & NCA \cite{dees2009analysis}, Si \cite{ding2009determination}, graphite \cite{tang2019systematic,chaouachi2021experimental}\\

Diffusion& PITT & NMC \cite{tsai2018single}, LFP \cite{han2004electrochemical}, LCO \cite{xia2006li}\\
coefficient $D$ & EIS  & NMC \cite{charbonneau2020impedance}, LCO \cite{tang2019systematic,xia2006li}, NCA \cite{dees2009analysis}, Si \cite{ding2009determination}\\
 & Depolarization & NMC \cite{amin2016characterization} \\
& Ab initio & NMC \cite{wei2015kinetics}, Si \cite{moon2016ab}, graphite \cite{zhong2017ab} \\
\midrule 
Partial molar & In situ X-ray & NMC \cite{Biasi2017}\\
volume $\Omega$ & Ab initio & NMC \cite{Woodcox2021}\\
\bottomrule 
\end{tabular}
\caption{Compilation of experimental and \textit{ab initio} studies characterising the concentration-dependent nature of the diffusion coefficient and the partial molar volume. The data is classified as a function of the material, spanning the most common electrode materials, the material property under consideration ($D$, $\Omega$), and the measurement technique.}
\label{table:1}
\end{table}

In this work, we conduct coupled deformation-diffusion finite element analyses of core-shell particles incorporating the role of concentration-dependent diffusion coefficient and partial molar volume. In addition, we examine the failure of core-shell structures (shell fracture and core-shell debonding), and build failure maps to guide design and structural integrity assessment. We particularise our analysis to a representative system - a spherical secondary particle with a $\mathrm{LiNi_{0.8}Mn_{0.1}Co_{0.1}O_{2}}$ (NMC811) core and $\mathrm{LiNi_{0.33}Mn_{0.33}Co_{0.33}O_{2}}$  (NMC111) shell, as illustrated in Fig. \ref{fig:cs}(b). We find that the consideration of concentration-dependent properties has a very significant impact on the results. A concentration-dependent partial molar volume brings dramatic differences in the stress distribution. Accounting for the sensitivity of diffusivity to Li content reveals a localisation phenomenon whereby a large concentration gradient is observed near the interface, which provides new insight into electrode underutilisation due to kinetic limitations. The failure maps show that previous models considering constant material properties may have been overly cautious in core radius design to avoid fracture.



\section{Methods}


In the following, we proceed to describe a fully coupled deformation-diffusion model, which accounts for both chemical strains and the effects of stress on diffusivity. The boundary conditions are discussed, and a physically rigorous model that assumes a continuous chemical potential across the core-shell interface, instead of concentration continuity, is presented. The parameters used in the model are listed, including constant parameters and concentration-dependent diffusion coefficient and partial molar volume. The theoretical model is numerically implemented by means of the finite element method using COMSOL Multiphysics.

\label{Sec:Theory}
\subsection{The coupled diffusion and stress model}
\label{Sebsec:model}
The lithium concentration in the particle, $c$, is governed by the conservation of species balance
\begin{equation}
    \frac{\partial c}{\partial t}+\bm{\nabla\cdot \mathrm{J}}=0, 
\end{equation}
\noindent where $t$ denotes time and $\bm{\mathrm{J}}$ is the concentration flux. 
The chemical potential $\mu$ can be expressed as \cite{wang2002effect}
\begin{equation}
    \mu=\mu_0+RT\ln(c)-\Omega\sigma_h,
    \label{eq:mu1}
\end{equation}
where $\mu_0$ is the reference chemical potential, \(R\) is the gas constant, \(T\) is the temperature, \(\Omega\) is the partial molar volume of lithium in the host material, and \(\sigma_h=\mathrm{tr}(\bm{\sigma})/3\) is the hydrostatic stress, with $\bm{\sigma}$ being the Cauchy stress tensor. Based on the kinetics theory, the diffusion flux is proportional to the gradient of chemical potential \cite{zhang2007numerical}
\begin{equation}
    \bm{\mathrm{J}}=-Mc\bm{\nabla} \mu,
    \label{eq:flux}
\end{equation}
where $M=D/RT$ is the mobility of lithium-ion in the host material and \(D\) is the diffusion coefficient. Combining Eqs. (\ref{eq:mu1}) and (\ref{eq:flux}), the species flux can be expressed as 
\begin{equation}
    \bm{\mathrm{J}}=-D\bm{\nabla} c+\frac{cD\Omega}{RT}\bm{\nabla}\sigma_h\quad \mathrm{and}\quad \bm{\mathrm{J}}\cdot\bm{\mathrm{n}}=J\quad \mathrm{at\ } \partial V,
    \label{eq:J}
\end{equation}
where \(J\) is the flux magnitude at the boundary, and $V$ is the volume of the particle. The driving force of lithium transport is the chemical potential which depends on the hydrostatic stress and lithium-ion concentration. The stress equilibrium requires that
\begin{equation}
    \bm{\nabla\cdot \mathrm{\sigma}}=\bm{0},
\end{equation}
The constitutive relation between the Cauchy stress tensor \(\bm{\sigma}\) and the strain tensor \(\bm{\varepsilon}\) is given by
\begin{equation}
\label{eq:ConstModel}
    \bm{\sigma}=\lambda\mathrm{tr}(\bm{\varepsilon-\varepsilon_{Li}})\bm{I}+2 G (\bm{\varepsilon-\varepsilon_{Li}}),
\end{equation}
where \(\lambda\) and \(G\) are the Lamé constants, and $\bm{I}$ is the identity matrix. In Eq. (\ref{eq:ConstModel}), \(\bm{\varepsilon_{Li}}\) is the chemical strain caused by lithium insertion which can be expressed as 
\begin{equation}
    \bm{\varepsilon_{Li}}=\frac{1}{3}\Omega\left(c-c_0\right)\bm{I},
\end{equation}
where \(c_0\) represents the initial lithium concentration in a stress-free state.

\subsection{Boundary conditions}
\label{Sebsec:bc}

Our numerical experiments employ a constant current constant voltage (CCCV) profile for the discharge (lithiation) process, and a constant current (CC) profile for the charge (delithiation) process of the cathode. A detailed definition and input parameters are given in \ref{sec:charge}. In the literature, a continuous concentration at the core-shell interface is typically assumed \cite{Hao2013}. However, this assumption cannot often be satisfied since the core and shell have different maximum concentrations. In this work, we use an arguably more physically rigorous model that assumes a continuous chemical potential \cite{Wu2017}, leading to the following relationship between the concentration of the two phases at the core-shell interface (see \ref{sec:ccp} for additional details);
\begin{equation}
    J_1(r=a)=J_2(r=a),
    \label{eq:J1equalJ2}
\end{equation}
\begin{equation}
    \mu_1(r=a)=\mu_2(r=a),
\end{equation}
\begin{equation}
    r=a:\quad c_1=U_{r e f, 1}^{-1}\left[\frac{\Omega_{2} \sigma_{h, 2}-\Omega_{1} \sigma_{h, 1}}{F}+U_{r e f, 2}\left(c_{2}\right)\right],
    \label{eq:c1}
\end{equation}

\noindent Here, $r$ is the radial position within a spherical particle. Also, in Eqs. (\ref{eq:J1equalJ2})-(\ref{eq:c1}) and hereafter, we use the numbers 1 and 2 to respectively represent the core and shell domains. In addition, $U_{r e f, 2}$ is the open circuit potential of the shell (see \ref{sec:ccp}). The other boundary and initial conditions are described as
\begin{equation}
    r=0:\quad \frac{\partial c_1}{\partial r}=0,
\end{equation}
\begin{equation}
    r=b:\quad J_2(r=b)=J_0,
\end{equation}
\begin{equation}
    t=0:\quad c_1=c_{1,0}\quad \mathrm{and}\quad c_2=c_{2,0},
\end{equation}
The stoichiometry $x$ and state of lithiation (SOL) are respectively defined as:
\begin{equation}
    x=\frac{c}{c_{max}},
\end{equation}
\begin{equation}
    \mathrm{SOL}=\frac{\int_{V}c\ \mathrm{d}V}{\int_{V}c_{max}\ \mathrm{d}V},
\end{equation}
where \(c_{max}\) is the maximum concentration of lithium in the host material. The stoichiometry is defined locally in the material while the SOL is defined globally at the particle or electrode level. $\mathrm{SOL}=0\%$ and $100\%$ respectively denote the electrode at full delithiation and full lithiation states. Based on the equations above, the concentration distribution and stresses inside the core-shell particle can be solved numerically using the finite element method. In addition, Zhao et al. \cite{Zhao2012} used dimensional analysis to determine the energy release rates of shell fracture $G_f$ and debonding $G_d$ as follows:
\begin{equation} \label{eq:Gf}
    G_{f}=Z \frac{\langle\bar{\sigma}_{\theta \theta}\rangle^{2}}{E_{2}}(b-a)
\end{equation}
\begin{equation} \label{eq:Gd}
    G_{d}=\pi \frac{\langle \sigma_{r r}^{c s}\rangle^{2}}{E_{e}}(b-a)
\end{equation}
where $\langle x \rangle = \left(x+|x|\right)/2$ are Macaulay brackets, $\left(b-a\right)$ represents the shell thickness, $\bar{\sigma}_{\theta \theta}=\left(2 \int_{a}^{b} \sigma_{\theta \theta} r \mathrm{~d} r\right) /\left(b^{2}-a^{2}\right)$ is the average hoop stress in the shell, $\sigma_{r r}^{c s}$ denotes the radial stress at the core-shell interface, $E_1$ and $E_2$ respectively denote Young's modulus of the core and shell, $E_e$ is the effective Young's modulus determined by $1/E_e=(1/E_1+1/E_2)/2$, and $Z$ represents a dimensionless parameter which equals 2 for a channel crack in a thin shell \cite{Zhao2012}. Eqs. (\ref{eq:Gf}) and (\ref{eq:Gd}) are derived upon the assumption of pre-existing defects of a length comparable to the thickness of the shell ($b-a$), as expected from processes like calendering \cite{Xu2023}.

\subsection{Constant and concentration-dependent parameters used in the model}
\label{Sebsec:parameter}
\begin{figure}[ht]
    \centering
    \includegraphics[width=1\textwidth]{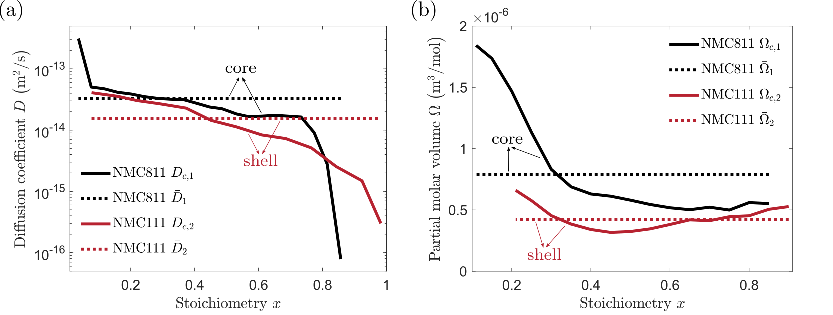}
    \caption{Concentration-dependent material properties: (a) diffusion coefficient $D_{c,1}$ (NMC811) and $D_{c,2}$ (NMC111), measured using GITT method \cite{Gaoy2020,Wu2012}.  (b) molar volume $\Omega_{c,1}$ (NMC811) and $\Omega_{c,2}$ (NMC111) derived from in situ X-ray diffraction data \cite{Biasi2017}. The constant effective values are shown using dotted lines.}
    \label{fig:DOmega}
\end{figure}

The particle is lithiated and delithiated between around $\mathrm{SOL}=20\%$ and 85$\%$ (see \ref{sec:charge} for details on the initial and final states). The magnitude of the constant current density employed is 100 $\mathrm{A/m^2}$; equivalent to 2.7 C or to a constant flux of $J=\mathrm{6.28\times 10^{-5}\ mol/(m^2\cdot s)}$. The constant parameters used in the finite element model are listed in Table \ref{table:2}. In this regard, it is worth emphasising that the effective partial molar volume $\bar{\Omega}$ and effective diffusion coefficient $\bar{D}$ are the average values of the concentration-dependent data obtained from experiments. Fig. \ref{fig:DOmega}(a) depicts the concentration-dependent diffusion coefficient $D_{c,1}$ and $D_{c,2}$ of the core (NMC811) and shell (NMC111), measured using GITT by Gao et al. \cite{Gaoy2020} and Wu et al. \cite{Wu2012} respectively. The GITT method is chosen here due to its accuracy and prevalence in the literature \cite{markevich2005comparison,barai2019comparison,wang2022unveiling}. As shown in experiments \cite{Gaoy2020,Wu2012}, a small hysteresis exists between discharge and charge processes. For the sake of simplicity, we base our analysis on the discharge data, which corresponds to the lithiation of the cathode. As shown in Fig. \ref{fig:DOmega}(a), both of the diffusion coefficient curves drop dramatically when the stoichiometry increases. In particular, the diffusion coefficient of NMC811 $D_{c,1}$ decreases by 3 orders of magnitude from $10^{-13}$ to $10^{-16}\ \mathrm{m^2/s}$. In terms of partial molar volume, De Biasi et al. \cite{Biasi2017} measured the volume change of NMC811 and NMC111 during charging using \textit{in situ} X-ray diffraction. The concentration-dependent partial molar volume $\Omega_c$ is derived from the volume change (see \ref{sec:omega1}) and shown in Fig. \ref{fig:DOmega}(b). The partial molar volume of NMC811 ($\Omega_{c,1}$) continues to decrease as the stoichiometry $x$ increases, while $\Omega_{c,2}$ of NMC111 decreases until $x$ reaches around 0.5 and then increases. A linear interpolation method is used with the data points provided in the literature. If the stoichiometry $x$ exceeds the limits shown in Fig. \ref{fig:DOmega}, the value of the relevant magnitude is assumed to remain constant thereafter. 
\begin{table}[ht]
\centering
\begin{tabular}{cccc}
\toprule  
Parameter&Symbol&Core (NMC811)&Shell (NMC111)\\
\midrule 
Core radius&$a$&4 µm&-\\
Shell thickness&$b-a$&-&1 µm\\
Maximum lithium concentration&$c_{max}$&51765 $\mathrm{mol/m^3}$ \cite{chen2020}&49000 $\mathrm{mol/m^3}$ \cite{ko2019porous}\\
Effective partial molar volume of lithium&$\bar{\Omega}$&$\mathrm{7.88\times 10^{-7}\ m^3/mol}$&$\mathrm{4.22\times 10^{-7}\ m^3/mol}$\\
Effective diffusion coefficient&$\bar{D}$&$\mathrm{3.26\times 10^{-14}\ m^2/s}$&$\mathrm{1.55\times 10^{-14}\ m^2/s}$\\
Young’s modulus&$E$&$\mathrm{184\ GPa}$ \cite{Woodcox2021}&$\mathrm{199\ GPa}$ \cite{Cheng2017}\\
Poisson’s ratio&$\nu$&$\mathrm{0.26}$ \cite{Woodcox2021}&$\mathrm{0.25}$ \cite{Cheng2017}\\
Flux at the outer surface&$J$&\multicolumn{2}{c}{$\mathrm{6.28\times 10^{-5}\ mol/(m^2\cdot s)}$}\\

\bottomrule 
\end{tabular}
\caption{Parameters used in the finite element model including the geometry of the core-shell structure, material properties (for the concentration-independent scenario), and magnitude of the constant flux.}
\label{table:2}
\end{table}

Three models are considered to separately investigate the effects of two concentration-dependent parameters: (1) the reference model, with constant diffusion coefficient $\bar{D}$ and partial molar volume $\bar{\Omega}$; (2) a model considering a concentration-dependent diffusion coefficient $D_c$ and constant partial molar volume $\bar{\Omega}$; and (3) a model considering a constant diffusion coefficient, $\bar{D}$, and a concentration-dependent partial molar volume $\Omega_c$.

\section{Results and discussion}
\label{Sec:Results}

The aforementioned coupled finite element model is used to gain insight into the behaviour of core-shell electrode particle configurations. We start, in Section \ref{subsec:DIS}, by characterising the concentration and stress distributions along the core-shell structure, comparing the outputs obtained with the models incorporating concentration-dependent diffusion coefficient ($D_c$) or partial molar volume ($\Omega_c$) to those of the reference model ($\bar{D}$, $\bar{\Omega}$). In particular, the notable and interesting effects of $\Omega_c$ on the stress distribution are thoroughly discussed and rationalised. Next, in Section \ref{Sec:Underutilisation}, we devote our efforts to understand a phenomenon revealed when considering a concentration-dependent diffusion coefficient: a significant underutilisation of cathode capacity at high SOL. Finally, in Section \ref{subsection:map} safe design maps are computed to design against fracture and debonding, considering both constant and concentration-dependent parameters.  \\

\subsection{Stress and Li concentration distributions in core-shell structures}
\label{subsec:DIS}

\begin{figure}[H]
    \centering
    \includegraphics[width=1\textwidth]{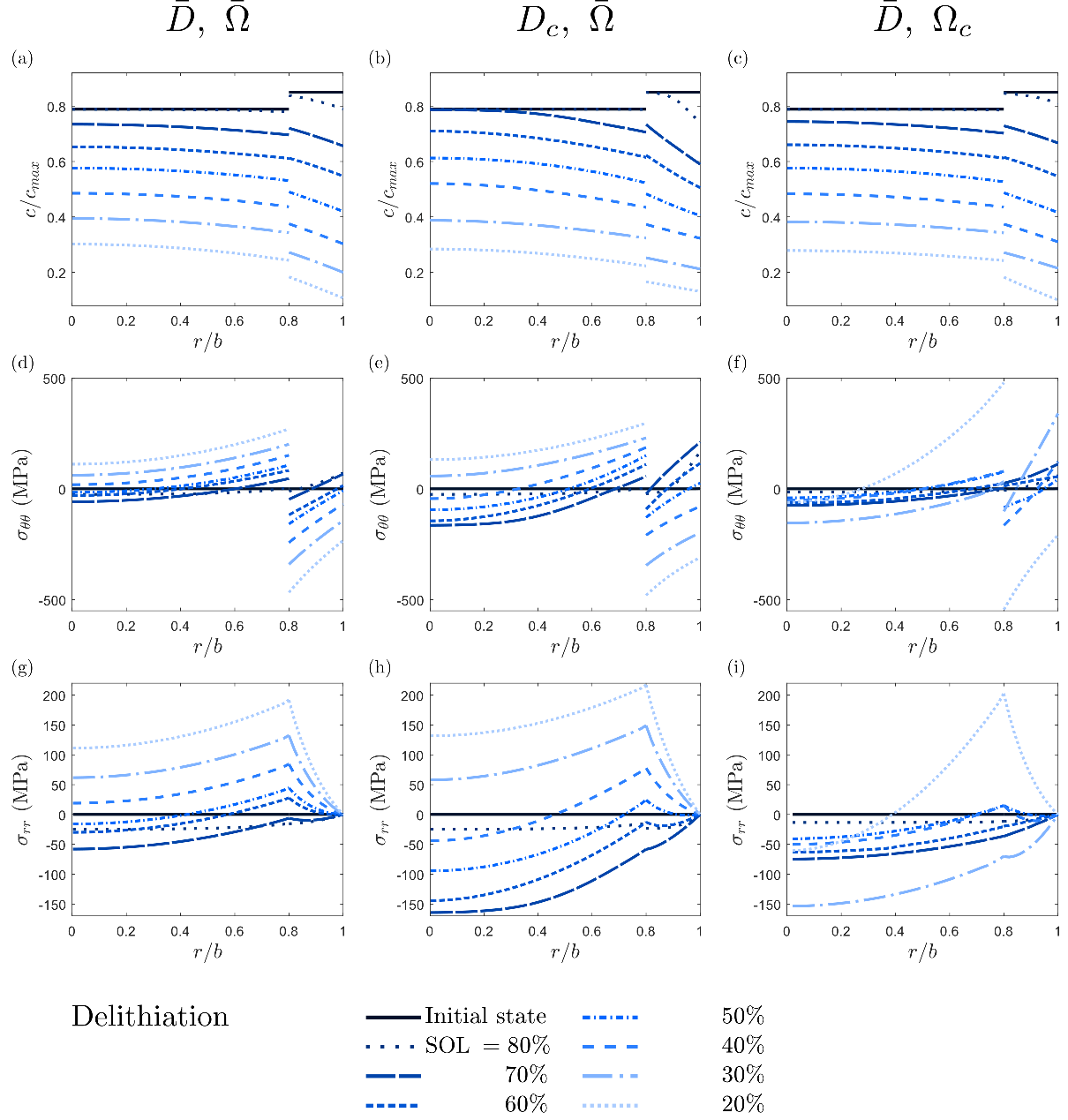}
    \caption{Delithiation results along the radial direction: dimensionless concentration for (a) $\bar{D}$, $\bar{\Omega}$, (b) $D_c$, $\bar{\Omega}$, (c) $\bar{D}$, $\Omega_c$; hoop stress for (d) $\bar{D}$, $\bar{\Omega}$, (e) $D_c$, $\bar{\Omega}$, (f) $\bar{D}$, $\Omega_c$; and radial stress for (g) $\bar{D}$, $\bar{\Omega}$, (h) $D_c$, $\bar{\Omega}$, (i) $\bar{D}$, $\Omega_c$. An overline denotes constant (averaged) properties while a $c$ subscript denotes concentration-dependent properties. The radial position $r$ is normalised by the outer radius $b$ of the core-shell system.}
    \label{fig:delith}
\end{figure}

\begin{figure}[H]
    \centering
    \includegraphics[width=1\textwidth]{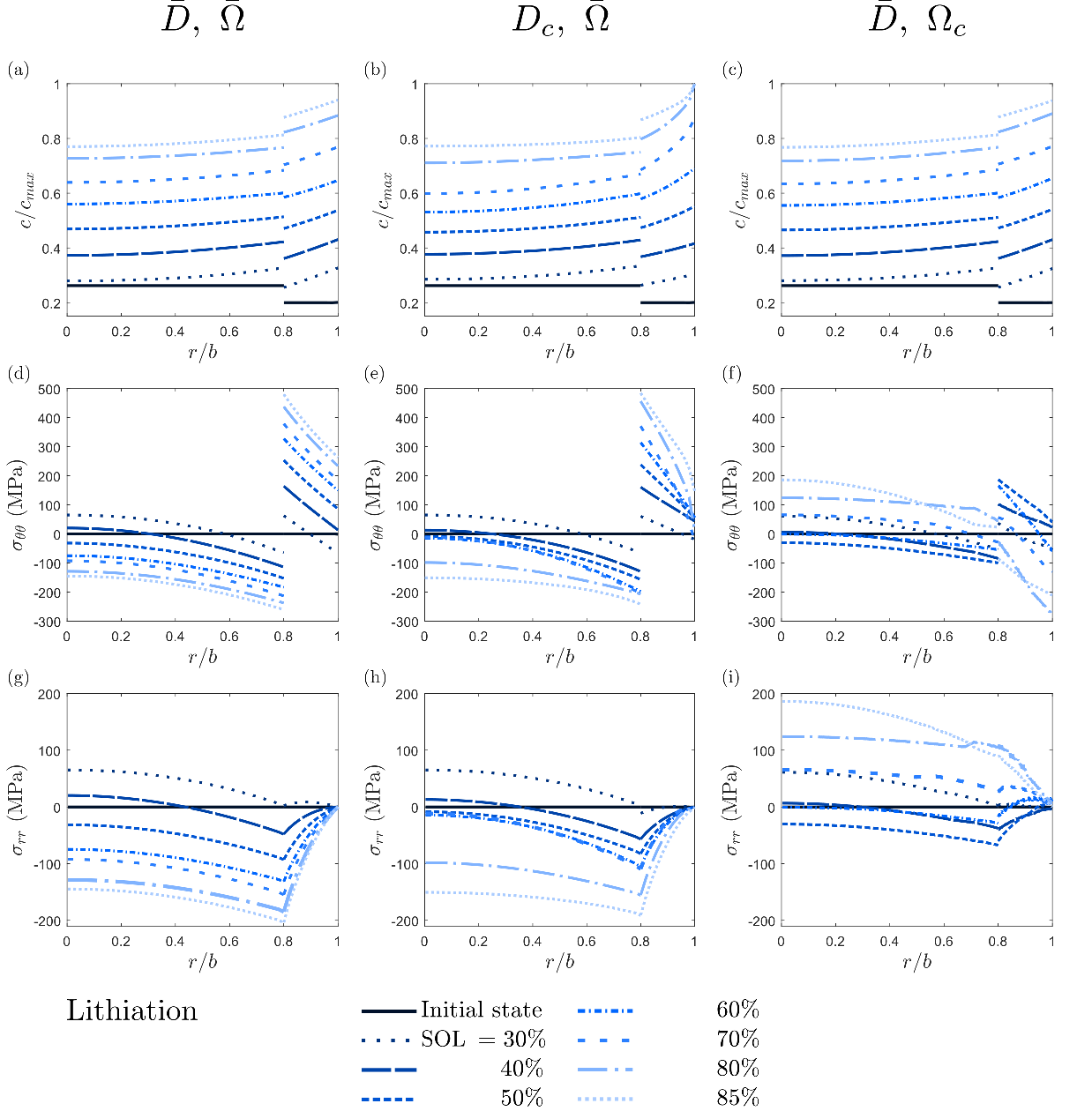}
    \caption{Lithiation results along the radial direction: dimensionless concentration for (a) $\bar{D}$, $\bar{\Omega}$, (b) $D_c$, $\bar{\Omega}$, (c) $\bar{D}$, $\Omega_c$; hoop stress for (d) $\bar{D}$, $\bar{\Omega}$, (e) $D_c$, $\bar{\Omega}$, (f) $\bar{D}$, $\Omega_c$; and radial stress for (g) $\bar{D}$, $\bar{\Omega}$, (h) $D_c$, $\bar{\Omega}$, (i) $\bar{D}$, $\Omega_c$. An overline denotes constant (averaged) properties while the subscript $c$ denotes concentration-dependent properties.}
    \label{fig:lith}
\end{figure}

Fig. \ref{fig:delith} depicts the radial distribution of lithium concentration and mechanical stress within the particle at different SOL levels during delithiation. The corresponding results during lithiation are displayed in Fig. \ref{fig:lith}. At the core-shell interface, located at $r/b=0.8$, concentration jumps are observed, as required to maintain a continuous chemical potential, see Figs. \ref{fig:delith}(a)-(c) and Figs. \ref{fig:lith}(a)-(c). The concentration profiles obtained with $D_c$ show slightly larger gradients at higher SOL; see Figs. \ref{fig:delith}(b) and \ref{fig:lith}(b), and compare to the results with $\bar{D}$ in Figs. \ref{fig:delith}(a) and \ref{fig:lith}(a). The larger gradient is a result of the drop in diffusivity at high lithium content, as demonstrated in Fig. \ref{fig:DOmega}(a). On the other hand, the concentration profiles are not visibly influenced by $\Omega_c$, as shown in Fig. \ref{fig:delith}(c) and Fig. \ref{fig:lith}(c). Although the partial molar volume amplifies the influence of hydrostatic stress on the diffusion flux, see Eq. (\ref{eq:J}), the concentration distribution at a certain SOL is dominated by the diffusion coefficient.

Figs. \ref{fig:delith}(d) and \ref{fig:delith}(g) and Figs. \ref{fig:lith}(d) and \ref{fig:lith}(g) show the evolution of DIS assuming constant material properties during delithiation and lithiation, respectively. These results can be rationalised as follows. When the partial molar volume is constant, the core and shell materials expand linearly with the amount of lithium inserted during intercalation (assuming a negligible difference in concentration gradient). The core expands more than the shell due to its higher partial molar volume, resulting in a compressive radial stress at the core-shell interface. The hoop stress in the core is also compressive because the shell limits the expansion of the core. In contrast, the shell experiences tensile hoop stress. As more lithium is inserted, the absolute values of DIS increase. A similar process occurs during delithiation, as illustrated in Fig. \ref{fig:failure}, but the tensile and compressive stresses are reversed. $D_c$ and $\bar{D}$ generate similar stresses at the same SOL due to their comparable concentration profiles and same constant partial molar volumes, as illustrated in Figs. \ref{fig:delith}(e)(h) and Figs. \ref{fig:lith}(e)(h), compared to the results with $\bar{D}$.\\ 

Incorporating the concentration-dependent partial molar volume has a significant impact on diffusion-induced stresses. During lithiation as shown in Fig. \ref{fig:lith}(i), the radial stress at the core-shell interface $\sigma_{rr}$ is no longer compressive but tensile once we reach SOL levels of 70\%. Fig. \ref{fig:DOmega}(b) shows that the partial molar volume of the core $\Omega_{c,1}$ decreases when the stoichiometry ($x$) increases, while the partial molar volume of the shell $\Omega_{c,2}$ decreases when $x$ increases up to around 0.5, and then increases. When considering $\Omega_c$ during lithiation, the shell expands more than with $\bar{\Omega}$ at higher SOL, while the core expands less than with $\bar{\Omega}$. When considering these effects, the radial stress at the core-shell interface becomes tensile for SOL higher than 60$\%$, as depicted in Fig. \ref{fig:lith}(i). Accordingly, the hoop stress in the shell becomes compressive as shown in Fig. \ref{fig:lith}(f). The results with a constant partial molar volume ($\bar{\Omega}$) show that the maximum tensile hoop stress in the shell, which may lead to shell fracture, occurs at $\mathrm{SOL}=85\%$. However, Fig. \ref{fig:lith}(i) with $\Omega_c$ shows a much smaller maximum tensile hoop stress in the shell, which occurs at $\mathrm{SOL}=50\%$ during lithiation. The maximum average hoop stress in the shell with $\Omega_c$ is around 3 times smaller than the result with $\bar{\Omega}$. A detailed quantitative analysis of the effects of $\Omega_c$ on DIS is included in \ref{sec:omega1}. The results show that considering the concentration-dependent partial molar volume, which captures the non-linear volume changes of the active materials, has a critical impact on DIS. In the NMC811@NMC111 core-shell structure, the tensile stresses that lead to shell failure could be overestimated by approximately three times if using a constant partial molar volume.

\subsection{Underutilisation of the electrode particle at a high state of lithiation}
\label{Sec:Underutilisation}

\begin{figure}[H]
    \centering
    \includegraphics[width=1\textwidth]{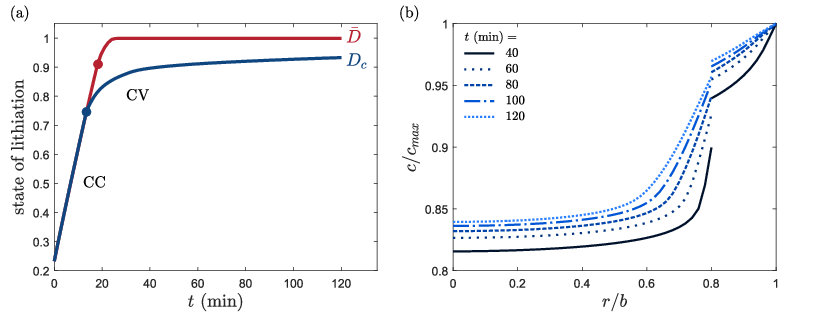}
    \caption{The results obtained with $D_c$ indicate that the particle is not fully utilised during discharge: (a) State of lithiation of the electrode particle as of function of time during lithiation. The red and blue curves represent the results obtained with $\bar{D}$ and $D_c$, respectively. The dots indicate the transition between CC (constant current) and CV (constant voltage) stages of the discharging process; (b) Radial distribution of dimensionless concentration at different times during lithiation with $D_c$. A significant concentration gradient is observed at the periphery of the core, caused by a low diffusion coefficient at high SOL.}
    \label{fig:cgradient}
\end{figure}
In the context of characterizing concentration fields during lithiation, an intriguing observation emerges: the SOL's growth rate significantly decelerates over time when considering a concentration-dependent diffusion coefficient ($D_c$), relative to a constant one ($\bar{D}$), once SOL surpasses $80\%$. This is primarily because $D_c$ experiences a dramatic reduction at high lithium content. To further investigate the intricacies of high SOL behavior, we extend our study by discharging the cathode particle until complete lithiation. The temporal evolution of SOL with $\bar{D}$ and $D_c$ is illustrated in Fig. \ref{fig:cgradient}(a). When assuming $\bar{D}$, the particle achieves complete lithiation rapidly during the CV (constant voltage) stage. However, when using $D_c$, the SOL remains at approximately $90\%$ for an extended duration of over 100 min during the CV stage, indicating significant barriers to full electrode utilisation due to reduced lithium mobility at high SOL. Fig. \ref{fig:cgradient}(b) shows the concentration distribution at various times near $\mathrm{SOL}=90\%$. Large concentration gradients can be observed in the core due to low diffusivity $D_c$. The diffusion coefficient of the NMC811 core drops by two orders of magnitude when the stoichiometry $x$ exceeds 0.75, as shown in Fig. \ref{fig:DOmega}(a). When the periphery of the core reaches a high lithium content, the diffusion coefficient reduces dramatically, hindering the diffusion within the core and generating a large concentration gradient. From the time $t=40$ min to $t=120$ min, the concentration level in the core grows very slowly, making it difficult for the particle to be fully lithiated. These findings demonstrate that the underutilisation of the cathode capacity can be attributed to kinetic limitations arising from low diffusion coefficient at high lithium concentration.\\
This phenomenon has been witnessed in experiments of single particles. Xu et al. \cite{Xu2022} observed that the centre of the NMC particle remained lithium deficient when SOL $>80\%$ during lithiation. A sharp lithium concentration gradient was found in the periphery of the particle. An additional CV step was applied for 2 hours, driving the the particle towards near-full lithiation state. These experimental results provide evidence of the inhomogenous concentration and capacity underutilisation in cathode particles, and are well aligned with our simulation results when considering a concentration-dependent diffusion coefficient, $D_c$. We emphasise that this phenomenon can not be captured if assuming a constant diffusion coefficient. Thus, our results highlight the importance of considering concentration-dependent diffusivity to investigate capacity loss due to kinetic limitations, especially in fast charging/discharging scenarios \cite{zhao2010fracture}.

\subsection{Design maps to avoid fracture and debonding}
\label{subsection:map}
\begin{figure}[H]
    \centering
    \includegraphics[width=1\textwidth]{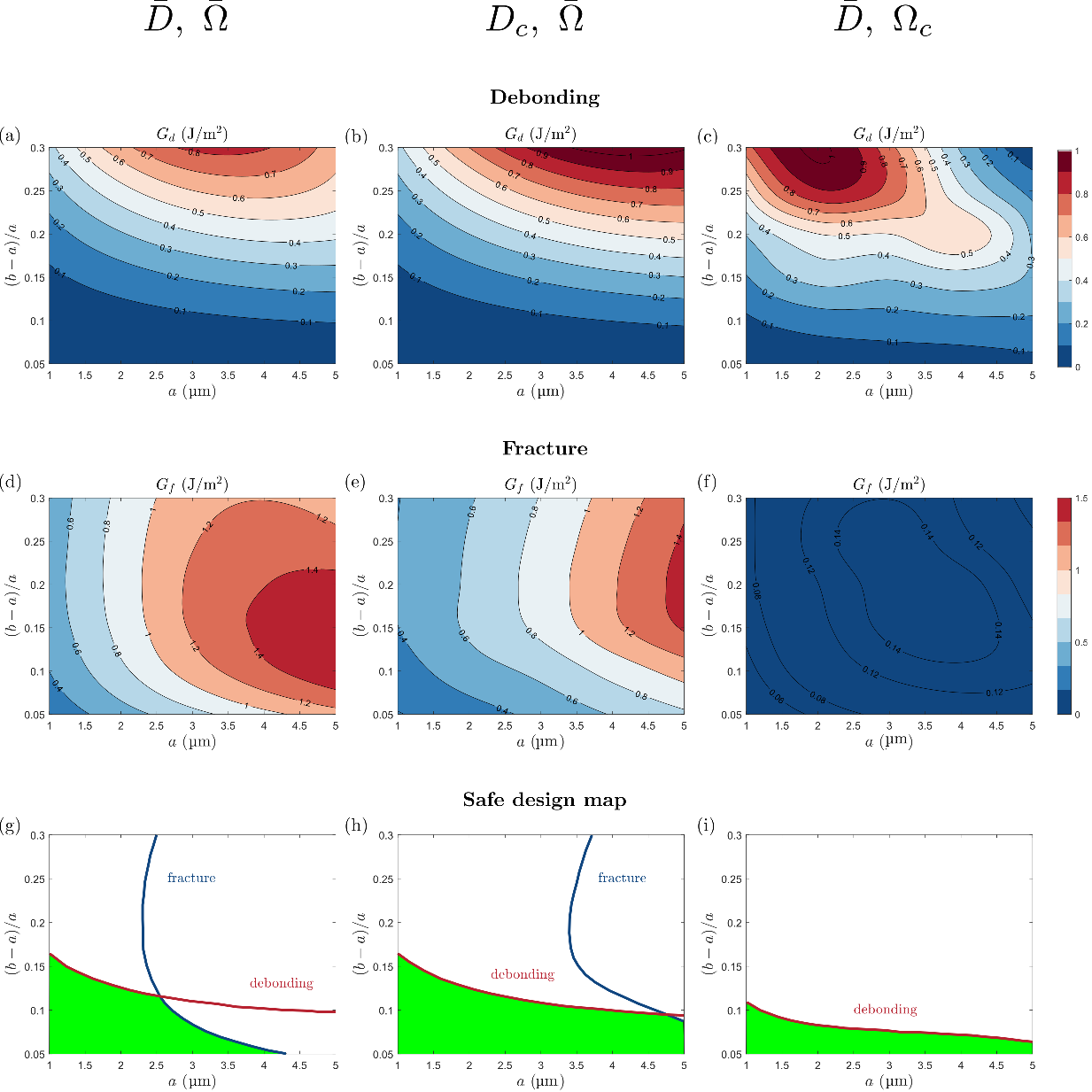}
    \caption{Design maps are developed to investigate the effects of core radius $a$ and relative shell thickness $(b-a)/a$ on debonding and fracture: the energy release rate of debonding $G_d$ for (a) $\bar{D}$, $\bar{\Omega}$, (b) $D_c$, $\bar{\Omega}$, (c) $\bar{D}$, $\Omega_c$; the energy release rate of fracture $G_f$  (d) $\bar{D}$, $\bar{\Omega}$, (e) $D_c$, $\bar{\Omega}$, (f) $\bar{D}$, $\Omega_c$; and the safe maps,  represented by green region, indicate the parameter combinations to prevent fracture and debonding assuming $G_d^c = 0.1\ \mathrm{J/m^2}$ and $G_f^c = 1\ \mathrm{J/m^2}$ for (g) $\bar{D}$, $\bar{\Omega}$, (h) $D_c$, $\bar{\Omega}$, (i) $\bar{D}$, $\Omega_c$.}
    \label{fig:contour}
\end{figure}

The energy release rates of shell fracture and debonding can be obtained based on the DIS and then used to guide the safe design of the core-shell structure. Higher energy release rates indicate a greater risk of fracture or debonding. To explore the safe design region, a parametric sweep is conducted to investigate the effect of core radius $a$ (ranging from 1 µm to 5 µm) and relative shell thickness $(b-a)/b$ (ranging from 0.05 to 0.3) under the lithiation/delithiation conditions described previously. These ranges of core radius and shell thickness are in agreement with typically reported values \cite{yoo2015novel,hou2017core,park2023core}. More details can be found in \ref{sec:charge}. Figs. \ref{fig:contour}(a)-(c) display the maximum energy release rate of debonding $G_d$ during delithiation up to $\mathrm{SOL}=20\%$. The simulations are carried out for $a=1,2,3,4,5$ µm, $(b-a)/a=0.05,0.1,0.15,0.2,0.25,0.3$ and cubic interpolation is used to provide contour maps. The results with two different models, ($\bar{D},\bar{\Omega}$) and ($D_c,\bar{\Omega}$), exhibit similar trends. The results indicate that thicker shells have a greater tendency for debonding. The results with ($\bar{D},\bar{\Omega}$) in Fig. \ref{fig:contour}(a) show that the most vulnerable region for debonding is thick shells with medium-sized cores with $a$ between 3 µm to 4 µm. On the other hand, for the ($D_c,\bar{\Omega}$) case study shown in Fig. \ref{fig:contour}(b), the most dangerous region is observed to be between 3.5 µm to 5 µm. These slight differences are caused by variations in concentration distribution in different geometries. As for DIS, the partial molar volume has significant effects on the debonding strength. The results with ($\bar{D},\Omega_c$) in Fig. \ref{fig:contour}(c) show similar contour values as the other two cases but different shapes, with the most dangerous area being for thick shells and small cores at around $a=2$ µm of size. \\

The maximum energy release rate of fracture $G_f$ during lithiation up to $\mathrm{SOL}=85\%$ is depicted in Figs. \ref{fig:contour}(d)-(f). For ($\bar{D},\bar{\Omega}$) and ($D_c,\bar{\Omega}$), $G_f$ reaches the maximum value at the end the lithiation. In general, larger particles with medium-sized shells are more prone to fracture. The maximum of $G_f$ for ($\bar{D},\Omega_c$) occurs between $\mathrm{SOL}=49\%$ to $56\%$, depending on the geometry. The maximum hoop stress in the shell happens in the middle of lithiation if we consider the non-linear expansion of the core and shell materials, as discussed in Section \ref{subsec:DIS}, indicating that this is the most risky moment for shell to fracture. Furthermore, $G_f$ with $\Omega_c$ is approximately 10 times smaller than with a constant $\bar{\Omega}$, predicting a significantly lower risk of fracture.\\

Shell debonding or fracture is predicted to occur if the maximum energy release rate of debonding or fracture exceeds the corresponding critical values. We choose the latter to be $G_f^c = 1\ \mathrm{J/m^2}$, which similar to that reported for NMC secondary particles \cite{sharma2023nanoindentation}, and the debonding fracture energy is taken to be one order of magnitude smaller, $G_d^c = 0.1\ \mathrm{J/m^2}$, to account for the relatively weaker bonding between the core and shell. Therefore, the region prone to fracture or debonding is illustrated in Figs. \ref{fig:contour}(g)-(i) with the green area indicating the safe design region. When assuming constant material properties, the safe region is limited by both the fracture and debonding, as illustrated in Fig. \ref{fig:contour}(g). Smaller cores with thinner shells are considered safe with $a$ limited to under 4.3 µm and $(b-a)/a$ under 0.16. When considering the concentration-dependent diffusion coefficient $D_c$, the safe region is dominated by debonding showing that thin shells are safe, as depicted in Fig. \ref{fig:contour}(h). The results with $\Omega_c$ in Fig. \ref{fig:contour}(i) indicate a smaller safe area of thin shells determined by debonding only. The relative shell thickness is limited to 0.11 at $a=1$ µm and 0.065 at $a=5$ µm. The results demonstrate that accounting for concentration-dependent material properties has a significant impact on fracture and debonding predictions. The safe design region determined with constant material properties is conservative in terms of core size.

\section{Conclusions}                                                                                             
\label{Sec:Concluding remarks}
In this study, we investigated the diffusion-induced stress (DIS) inside a spherical particle with an NMC811@NMC111 core-shell structure during lithiation and delithiation. We obtained safe design maps to prevent shell debonding and fracture, with a discussion on the concentration-dependent diffusion coefficient and partial molar volume. The main findings are 
\begin{itemize}
  \item The concentration-dependent partial molar volume had minimal impact on the concentration distribution at the same SOL level but significantly affected the DIS. The tensile stresses that lead to shell failure could be overestimated by approximately three times if using constant partial molar volume;
  \item Underutilisation of the electrode capacity is predicted when considering concentration-dependent diffusion coefficient, due to reduced lithium mobility at high state of lithiation. This finding can provide an explanation for reduced accessible capacity of electrode, as observed in experiments;
  \item In terms of safe design maps, the results with constant material properties suggest smaller ($a\lesssim4.3$ µm) particles to avoid shell fracture and thinner coatings ($(b-a)/a\lesssim0.165$) to avoid shell debonding. However, when considering concentration-dependent diffusion coefficient or partial molar volume, the core size is no longer the determining factor within the range of $a\leq5$ µm. This indicates that previous models that consider constant material properties might have been conservative in designing the core radius to avoid fracture.
\end{itemize}
These results demonstrate the importance of considering concentration-dependent material properties in modelling electrode core-shell particles of lithium-ion batteries. More research efforts are needed to characterize the concentration-dependent transport, thermodynamic, and mechanical properties of different active materials. Similar studies are also needed to investigate the effects of temperature on the material properties. In addition, the present analysis could be extended to the electrode level (to capture spatial changes in mechanical and electrochemical behaviour \cite{Boyce2022,Boyce2024}) and to accommodate realistic particle morphologies \cite{Ai2022,Parks2023}.

\section*{Acknowledgements}
E. Mart\'{\i}nez-Pa\~neda acknowledges financial support from UKRI's Future Leaders Fellowship programme [grant MR/V024124/1]. W. Ai acknowledges the financial support from the Natural Science Foundation of Jiangsu Province, China (grant number BK20220795).




\appendix

\section{Charging/discharging conditions}
\label{sec:charge}
Table \ref{table:3} provides charging (delithiation) and discharging (lithiation) conditions for the cathode particle. The initial and final states are specified, with lithiation following a constant current constant voltage (CCCV) profile and delithiation following a constant current (CC) profile.
\begin{table}[H]

\begin{tabular}{cccc}

\toprule  
& Lithiation & Delithiation\\
\midrule 
& $c_2=0.2c_{max,2}$,   & $c_2=0.85c_{max,2}$,  \\
Initial state  & $c_1$ determined by the interface condition. & $c_1$ determined by the interface condition.\\
 & $\mathrm{SOL}=23\%$ & $\mathrm{SOL}=82\%$\\
 \hline
 & \textbf{CC} & \textbf{CC} \\
  & A constant flux is applied until $c_2$ & A constant flux is applied until $c_2$\\
Stage 1  & reaches $c_{max,2}$ at the outer surface. & reduces to 0 at the outer surface. \\
 & The magnitude of the current density is  & The magnitude of the current density is \\
 & $i=100\ \mathrm{A/m^2}$, equivalent to 2.7 C. &$i=100\ \mathrm{A/m^2}$, equivalent to 2.7 C. \\
 \hline
Stage 2 & \textbf{CV} & none\\
 & $c_2=c_{max,2}$ at the outer surface &  \\
\hline 
Final state & $\mathrm{SOL}=85\%$ & $\mathrm{SOL}=20\%$\\
\bottomrule

\end{tabular}

\caption{CCCV lithiation and CC delithiation conditions.}
\label{table:3}
\end{table}

The constant flux at the outer surface reads, 
\begin{equation}
    J=\frac{i}{Fa_sL}
\end{equation}
\noindent Here, $L$ refers to the thickness of the electrode and $a_s$ denotes the active surface area of the electrode materials per unit volume. For spherical particles, the active surface area can be expressed as $a_s=2\varepsilon_s/b$ \cite{Wu2017}, where $\varepsilon_s$ represents the volume fraction of active materials. In our numerical simulations, we consider $L=50$ µm and $\varepsilon_s=0.55$, which yields $J=\mathrm{6.28\times 10^{-5}\ mol/(m^2\cdot s)}$ for a current density of $i=100\ \mathrm{A/m^2}$.

\section{Continuous chemical potential}
\label{sec:ccp}
\begin{figure}[H]
    \centering
    \includegraphics[width=10cm]{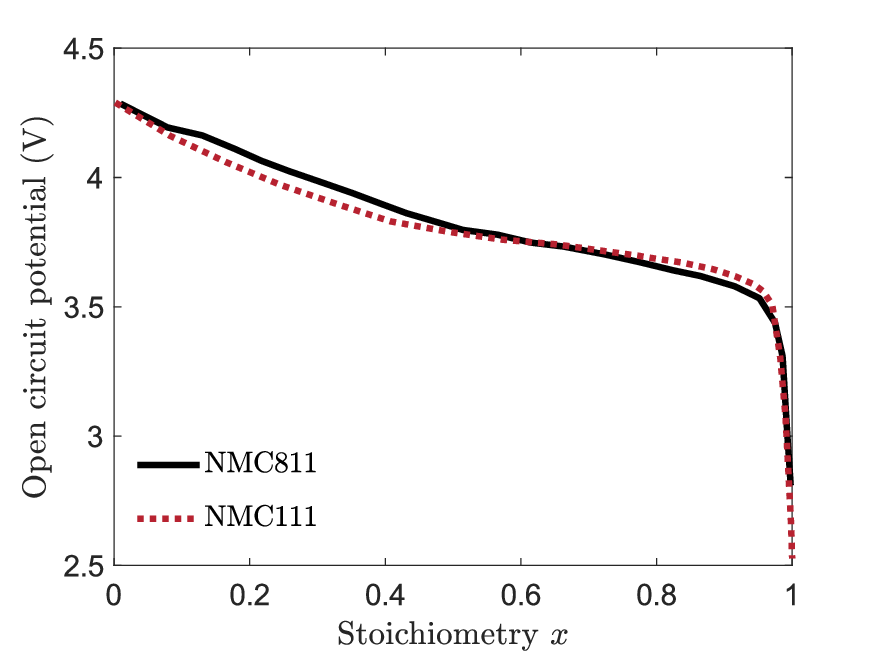}
    \caption{OCP profile of the NMC811 core and NMC111 shell materials. \cite{Wu2017,Wu2012}}
    \label{fig:ocp}
\end{figure}

Here, we establish the relationship between $c_1$ and $c_2$ at the core-shell interface, assuming a continuous chemical potential. This follows closely the approach outlined in Ref. \cite{Wu2017}. 
We start by defining the chemical potential of lithium in active particles as $\mu$
\begin{equation}
    \mu=\mu_{\mathrm{Li}-\Theta}-\mu_{\Theta},
\end{equation}
where $\mu_{\mathrm{Li}-\Theta}$ is the chemical potential of lithium in the lattice, and $\mu_{\Theta}$ is the chemical potential of vacancy in the lattice. The chemical potentials are determined as follows:
\begin{equation}
\mu_{\mathrm{Li}-\Theta}=\mu_{\mathrm{Li}-\Theta}^{0}+R T \ln a_{\mathrm{Li}-\Theta}-\Omega_{\mathrm{Li}-\Theta} \sigma_{h},
\end{equation}
\begin{equation}
\mu_{\Theta}=\mu_{\Theta}^{0}+R T \ln a_{\Theta}-\Omega_{\Theta}  \sigma_{h},
\end{equation}
Here, we have $a_{j}$ denoting the activity of phase $j$, $\mu_{j}^{0}$ signifying the standard state chemical potential of phase $j$, $\Omega_{j}$ indicating the partial molar volume of phase $j$, and $\sigma_{h}$ representing the hydrostatic stress in the lattice.\\
Thus, 
\begin{equation}
    \mu=\mu_{\mathrm{Li}-\Theta}^{0}-\mu_{\Theta}^{0}+R T \ln \left( \frac{a_{\mathrm{Li}-\Theta}}{a_{\Theta}}\right)-\Omega \sigma_{h},
    \label{eq:mu}
\end{equation}
where $\Omega=\Omega_{\mathrm{Li}-\Theta}-\Omega_{\Theta}$ is the partial molar volume of lithium in the solid, $\mu_{\Theta}^{0}$ and $\mu_{L i-\Theta}^{0}$ are both constants. The measured open circuit potential $U_{r e f}$ is determined by the Nernst equation as follows
\begin{equation}
    U_{ref}=U_{ref}^{0}+\frac{R T}{F} \ln \left(\frac{a_{\Theta}}{a_{\mathrm{Li}-\Theta}}\right),
\end{equation}
where $U_{r e f}^{0}=\left(\mu_{\mathrm{Li}}^{0}+\mu_{\Theta}^{0}-\mu_{\mathrm{L}-\Theta}^{0}\right) / F$, $F$ is the Faraday constant, $\mu_{\mathrm{Li}}^{0}$ is the chemical potential of lithium.
Thus, Eq. (\ref{eq:mu}) becomes
\begin{equation}
    \mu=-F U_{r e f}+\mu_{\mathrm{Li}}^{0}-\Omega \sigma_{h},
\end{equation}
The continuity of chemical potential $\mu_1=\mu_2$ is transformed to
\begin{equation}
    -F U_{r e f, 1}\left(c_{1}\right)-\Omega_{1} \sigma_{h, 1}=-F U_{r e f, 2}\left(c_{2}\right)-\Omega_{2} \sigma_{h, 2},
\end{equation}
Thus, the lithium concentration of the core at the interface can be expressed as 
\begin{equation}
    c_1=U_{r e f, 1}^{-1}\left[\frac{\Omega_{2} \sigma_{h, 2}-\Omega_{1} \sigma_{h, 1}}{F}+U_{r e f, 2}\left(c_{2}\right)\right],
\end{equation}
where $U_{r e f, 1}^{-1}$ is the inverse function of the open circuit potential of the shell. The measured open circuit potential profiles of the core and shell materials \cite{Wu2017,Wu2012} are shown in Fig. \ref{fig:ocp}. 

\section{Effects of $\Omega_c$ on DIS}
\label{sec:omega1}
First, we explain how $\Omega_c$ is derived from unit volume change obtained using X-ray experiments. Then, we quantify the effects of $\Omega_c$ on DIS, compared to results with constant partial molar volume, through theoretical analysis. The partial molar volume is the volume change per mole of lithium when an additional mole of lithium is added to the active material:
\begin{equation}
   \Omega  =\frac{3\varepsilon}{\Delta c}=\frac{3}{\Delta c}\left(\left(1+\frac{\Delta V}{V_0}\right)^{\frac{1}{3}}-1\right) =\frac{3}{C_{max}\Delta x}\left(\left(1+\frac{\Delta V}{V_0}\right)^{\frac{1}{3}}-1\right), 
\end{equation}
where $\Delta c$ is the variation of lithium concentration, $C_{max}$ is the maximum concentration, $\Delta x$ is the variation of stoichiometry, $\Delta V$ is the volume change, and $V_0$ is the initial volume. Thus, as shown in Fig. \ref{fig:DOmega}(b), the concentration-dependent partial molar volume can be obtained from the unit volume change \cite{Biasi2017}.\\
To better understand the shift in stress states during lithiation when accounting for $\Omega_c$, from compressive to tensile for $\sigma_{rr}^{cs}$ or from tensile to compressive for $\bar{\sigma}_{\theta \theta}$ in the shell, we examine the analytical solutions of the stresses, which are in good agreement with the finite element results. Eq. (\ref{eq:sr}) and Eq. (\ref{eq:st}) are the analytical expressions of $\sigma_{rr}^{cs}$ and $\sigma_{\theta \theta}$ respectively. Similarly, Eq. (\ref{eq:src}) and Eq. (\ref{eq:stc}) are the corresponding analytical expressions when accounting for $\Omega_c$. 
\begin{equation}
\sigma_{r r}^{c s}=\frac{2 E_1 E_2}{a^3} \frac{\overbrace{\int_a^b \left(c_2 -c_{0,2}\right)\bar{\Omega}_2\ r^2 \mathrm{~d} r}^\text{$\Phi_1$}-\overbrace{\left[\left(\frac{b}{a}\right)^3-1\right]  \int_0^a \left(c_1 -c_{0,1}\right) \bar{\Omega}_1\ r^2 \mathrm{~d} r}^\text{$\Phi_2$}}{\left(\frac{b}{a}\right)^3\left[E_1\left(1+\nu_2\right)+2 E_2\left(1-2 \nu_1\right)\right]+2\left[E_1\left(1-2 \nu_2\right)-E_2\left(1-2 \nu_1\right)\right]},
\label{eq:sr}
\end{equation}
\begin{equation}
\sigma_{r r}^{c s}=\frac{2 E_1 E_2}{a^3} \frac{\overbrace{\int_a^b \left(c_2 -c_{0,2}\right)\Omega_{c,2}\ r^2 \mathrm{~d} r}^\text{$\Phi_1$}-\overbrace{\left[\left(\frac{b}{a}\right)^3-1\right]  \int_0^a \left(c_1 -c_{0,1}\right) \Omega_{c,1}\ r^2 \mathrm{~d} r}^\text{$\Phi_2$}}{\left(\frac{b}{a}\right)^3\left[E_1\left(1+\nu_2\right)+2 E_2\left(1-2 \nu_1\right)\right]+2\left[E_1\left(1-2 \nu_2\right)-E_2\left(1-2 \nu_1\right)\right]},
\label{eq:src}
\end{equation}

\begin{equation}
\begin{split}
\sigma_{\theta \theta}=&\overbrace{\left. -\frac{a^3}{b^3-a^3}\left[1+\frac{1}{2}\left(\frac{b}{r}\right)^3\right] \sigma_{r r}^{c s}\right \}}^\text{$\Phi_3$}\\
&\underbrace{+\frac{ E_2}{3\left(1-\nu_2\right)}\left[\frac{1}{b^3-a^3}\left(2+\left(\frac{a}{r}\right)^3\right) \int_a^b \left(c_2 -c_{0,2}\right)\bar{\Omega}_2  r^2 \mathrm{~d} r 
+\frac{1}{r^3} \int_a^r \left(c_2 -c_{0,2}\right)\bar{\Omega}_2  r^2 \mathrm{~d} r-\left(c_2 -c_{0,2}\right)\bar{\Omega}_2 \right]}_\text{$\Phi_4$} 
\end{split}
\label{eq:st}
\end{equation}

\begin{equation}
\begin{split}
\sigma_{\theta \theta}=&\overbrace{\left. -\frac{a^3}{b^3-a^3}\left[1+\frac{1}{2}\left(\frac{b}{r}\right)^3\right] \sigma_{r r}^{c s}\right \}}^\text{$\Phi_3$}\\
&+\frac{ E_2}{3\left(1-\nu_2\right)}\left[\frac{1}{b^3-a^3}\left(2+\left(\frac{a}{r}\right)^3\right) \int_a^b \left(c_2 -c_{0,2}\right)\Omega_{c,2}  r^2 \mathrm{~d} r \right]\\
&\underbrace{+\frac{ E_2}{3\left(1-\nu_2\right)}\left[ \frac{1}{r^3} \int_a^r \left(c_2 -c_{0,2}\right)\Omega_{c,2}  r^2 \mathrm{~d} r\ -\ \left(c_2 -c_{0,2}\right)\Omega_{c,2} \right]}_\text{$\Phi_4$}    
\end{split}
\label{eq:stc}
\end{equation}

\begin{figure}[H]
    \centering
    \includegraphics[width=0.8\textwidth]{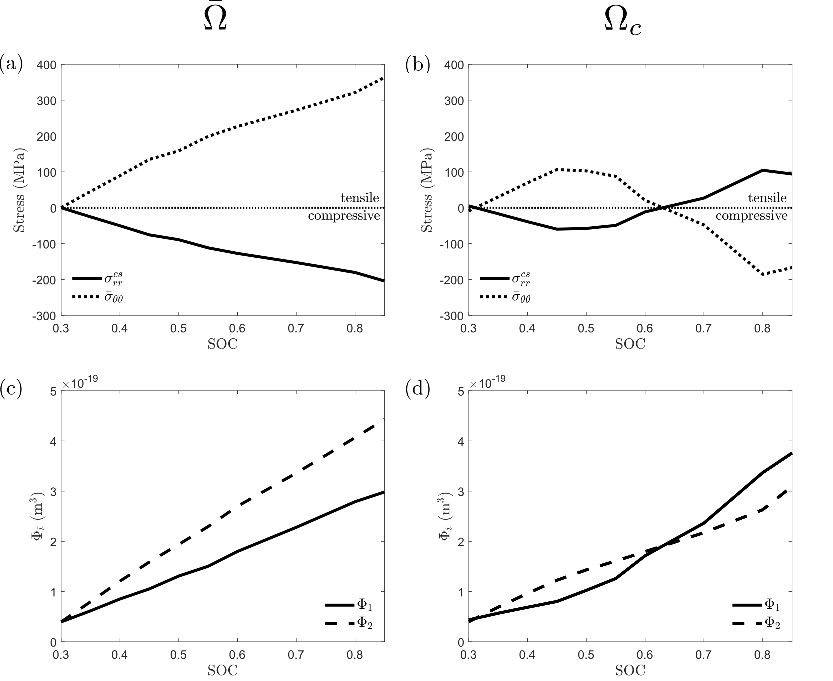}
    \caption{The evolution of radial stress at the core-shell interface $\sigma_{rr}^{cs}$ and average hoop stress in the shell $\bar{\sigma}_{\theta\theta}^s$ during lithiation for (a) $\bar{\Omega}$ and (b) $\Omega_c$; The evolution of $\Phi_1$ and $\Phi_2$ during lithiaiton for (c) $\bar{\Omega}$ and (d) $\Omega_c$.}
    \label{fig:stress}
\end{figure}

We can ignore $\Phi_4$ in $\sigma_{\theta \theta}$ and consider that $\sigma_{rr}^{cs}$ and $\sigma_{\theta \theta}$ have opposite signs cause numerically $| \Phi_4 / \Phi_3 | \ll 1$. When $\Phi_1$ is smaller than $\Phi_2$, $\sigma_{rr}^{cs}$ is compressive. When $\Phi_1$ is larger than $\Phi_2$, $\sigma_{rr}^{cs}$ is tensile. Fig. \ref{fig:stress} shows the evolution of $\sigma_{rr}^{cs}$,  $\bar{\sigma}_{\theta\theta}$, $\Phi_1$, and $\Phi_2$ in relation to SOL during lithiation, as obtained from finite element simulations. Figs. \ref{fig:stress}(a)(c) represent the evolution considering constant $\bar{\Omega}$, while Figs. \ref{fig:stress}(b)(d) show the evolution accounting for concentration-dependent $\Omega_c$. With $\bar{\Omega}$, $\Phi_1$ and $\Phi_2$ exhibit almost linear growth with the SOL, while $\sigma_{rr}^{cs}$ remains compressive during the lithiaiton process. However, it is worth noting that $\Phi_1$ grows faster than $\Phi_2$ when considering $\Omega_c$. This is because $\Omega_{c,1}$ reduces when the stoichiometry $x$ increases while $\Omega_{c,2}$ decreases until $x$ reaches approximately 0.5 then increases. As a result, $\sigma_{rr}^{cs}$ transforms from compressive to tensile during lithiation when considering $\Omega_c$. Incorporating concentration-dependent partial molar volume has significant impacts on the DIS results.

\end{document}